\documentstyle[]{mn}

\newcommand{\ipp}{\mbox{IP~Peg}} 
\newcommand{\etal}{\mbox{et\ al.\ }}
\newcommand{\kmsec}{\,\mbox{$\mbox{km}\,\mbox{s}^{-1}$}}

\newcommand{\hb}{\hbox{$\hbox{H}\beta$}}
\newcommand{\hg}{\hbox{$\hbox{H}\gamma$}}

\newcommand{\hei}{\hbox{$\hbox{He\,{\sc i}\,$\lambda$4472\,\AA}$}}
\newcommand{\heii}{\hbox{$\hbox{He\,{\sc ii}\,$\lambda$4686\,\AA}$}}

\newcommand{\mgii}{\hbox{$\hbox{Mg\,{\sc ii}\,$\lambda$4481\,\AA}$}}

\title[Spiral structure in \ipp]
      {Spiral structure in IP~Pegasi; how persistent is it?}
\author[L.\ Morales-Rueda, T.\, R.\ Marsh, I.\ Billington]
       {L.\ Morales-Rueda$^1$, T.\, R.\ Marsh$^1$ and I.\ Billington$^2$ \\
       $^1$Dept of Physics and Astronomy, Southampton University, 
       (lmr@astro.soton.ac.uk, trm@astro.soton.ac.uk)\\
       $^2$Swiss Bank Corp., City of London}
\date{Accepted ...
      Received ...;
      in original form ...}

\begin{document}

\maketitle

\begin{abstract}
  We present spectroscopy of the dwarf nova IP~Pegasi taken during two
  consecutive nights, 5 and 6 days after the start of an outburst.
  Even this late in the outburst, Doppler maps show marked spiral
  structure in the accretion disc, at least as strongly as seen
  earlier in other outbursts of \ipp. The spiral shocks are present on
  both nights with no diminution in strength from one night to the
  next.  The light curves of the lines show an offset to earlier
  phases, with the mid-eclipse of the emission lines displaced to
  phases between $-0.015\pm0.001$ and $-0.045\pm0.009$. This cannot be
  explained by the presence of the accretion shocks.  As well as the
  fixed spiral pattern, the disc shows strong flaring in the Balmer
  and \heii\ lines.  Irradiation-induced emission is seen from the
  companion star in the Balmer, He\,{\sc i}, He\,{\sc ii}, Mg\,{\sc
    ii}, C\,{\sc ii}, and other lines.  The emission is located near
  the poles of the companion star, suggesting that the accretion disc
  shields the companion star substantially and thus has an effective
  H/R of order 0.2 at EUV wavelengths. The Balmer emission is
  distinctly broader than the other lines consistent with non-Doppler
  broadening.
\end{abstract}
 
\begin{keywords}
  
accretion, accretion   discs  --   binaries:  spectroscopic  --  line:
profiles -- stars: mass-loss -- stars: novae, cataclysmic variables --
stars: individual: \ipp.
 
\end{keywords}

\section{Introduction}

%Cataclysmic variables (CVs) are semi-detached binary systems in which
%one of the components, a main sequence star, fills its Roche lobe and
%transfers mass to the other component, a white dwarf.
\ipp\ is an eclipsing cataclysmic variable (CV) of period 3.8 hours
(Wolf \etal 1993) that shows dwarf nova outbursts approximately every
3 months. During these high states, which last about 12 days, the
system goes from V magnitude 15 to 12. \ipp\ has received much
publicity recently as it was found to show strong spiral structure in
the accretion disc during outburst (Steeghs, Harlaftis \& Horne 1997;
Harlaftis \etal 1999).  Spiral shocks generated by tidal forces from
the companion star, were proposed as a possible mechanism of angular
momentum transport by Sawada, Matsuda \& Hachisu (1986) and Spruit
\etal (1987) , but had never been observed until Steeghs et al.'s
work.  The spiral pattern seems to be present only during outburst,
consistent with the idea that, to avoid producing too tightly wound a
spiral, the gas in the disc must have low Mach number (i.e. the speed
of the flow is not very large compared with the speed of sound in the
disc).  Two other systems that possibly show signs of spiral shocks in
their accretion discs are SS~Cyg (Steeghs \etal 1996) and EX~Dra
(Steeghs priv. comm.).

In the standard disc instability model of dwarf nova outbursts, the
viscosity in the disc increases significantly as the disc makes the
transition to outburst. This will drive the disc outwards against the
tidal field of the companion star which truncates the disc. It is at
this point that one expects the spiral shocks to develop. However, it
is not known yet how the spiral structure evolves with time or if it
is present in every outburst, therefore observations of the outbursts
at all epochs are of interest. The detections by Steeghs \etal (1997)
and Harlaftis \etal (1999) were made $1.5$ and 3 days after the start
of the August 1993 and the November 1996 outbursts respectively,
whereas spectroscopic data taken 8 days after the start of the August
1993 outburst by Steeghs \etal (1996) only hints at their presence.
In this paper, we present data taken on two nights, 5 and 6 days after
\ipp\ went into outburst in August 1994. This covers times
intermediate between phases when the spiral structure is seen
strongly, and when it is difficult to detect. We report the presence
of strong spiral structure in the accretion disc on both nights.  We
also detect the effect of spiral structure upon the emission line
eclipse light curves, and shielding of the companion star by the disc.

We start by describing the observations, and then present our results
and conclusions.

\section{Observations}

\begin{figure}
\begin{picture}(100,0)(-270,250)
\put(0,0){\includegraphics{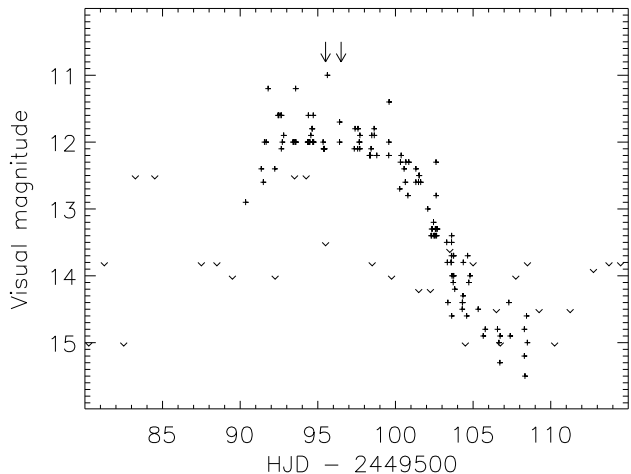}}
\noindent
\end{picture}
\vspace{75mm}
\caption{The visual light curve of \ipp\ during the August 1994 outburst
  obtained from the American Association of Variable Star Observers
  (http://www.aavso.org/). The arrows indicate the times of our
  spectrophotometric observations.}
\label{obs:aavso}
\end{figure}

We obtained spectrophotometry of \ipp\ on the nights of 1994 August 30
and 31, 5 and 6 nights after the start of an outburst, using the
Intermediate Dispersion Spectrograph mounted on the 2.5\,m Isaac
Newton Telescope (INT) on La Palma. Table~\ref{obs:jo} gives a journal
of observations; Fig.~\ref{obs:aavso} shows a visual light curve
obtained by the American Association of Variable Star Observers
(AAVSO) during the August 1994 outburst, with arrows indicating the
days on which we made observations.  The standard readout mode was
used in conjunction with a Tektronix CCD windowed to
1024\,$\times$\,90 pixels to reduce dead time. Exposure times were
generally 150\,s, with $\sim$\,70\,s dead time, and the spectral
resolution corresponds to $100\,\kmsec$ at \hb.  The 1200
lines\,mm$^{-1}$ grating R1200Y was used to cover the wavelength range
$\lambda\lambda$4040 -- 4983\AA\ enabling simultaneous monitoring of
\hb, \hg\ and \heii. A total of 164 usable spectra were obtained.

After debiasing and flat-fielding the frames by tungsten lamp
exposures, spectral extraction proceeded according to the optimal
algorithm of Horne (1986).  The data were wavelength calibrated using
a CuAr arc lamp and corrected for instrumental response and extinction
using the flux standard HD\,19445 (Oke \& Gunn 1983).  The
spectrograph slit orientation allowed a nearby star to be employed as
calibration for light losses on the slit.

\begin{table}
\centering
\begin{minipage}{84mm}
\caption{Journal of observations. $E$ is the cycle number plus binary 
  phase with respect to the ephemeris given by Wolf \etal (1993).
  Phases have been adjusted by 0.043 so that phase 0 corresponds to
  superior conjunction of the white dwarf.}
\label{obs:jo}
\begin{center}
\begin{tabular}{cccccr}
\multicolumn{1}{c}{Date} & \multicolumn{1}{c}{Start} & 
\multicolumn{1}{c}{End} & \multicolumn{1}{c}{Start} & 
\multicolumn{1}{c}{End} & \multicolumn{1}{c}{No. of} \\
\multicolumn{1}{c}{(1994 Aug)} & \multicolumn{2}{c}{(UT)} & 
\multicolumn{2}{c}{($E -$ 2\,500)} & \multicolumn{1}{c}{spectra} \\ \hline
30 & 1.48  & 5.99 & 158.004 & 159.193 & 69 \\
31 & 23.66 & 6.08 & 163.844 & 165.536 & 95 \\
\end{tabular}
\end{center}
\end{minipage}
\end{table}

\section{Results}
\subsection{Average spectra}

\begin{figure}
\begin{picture}(100,0)(-270,250)
%\put(0,0){\special{psfile="ippavspe4v2.ps" 
%hoffset=-8 voffset=32 angle=90 vscale=37 hscale=37}}
\put(0,0){\includegraphics{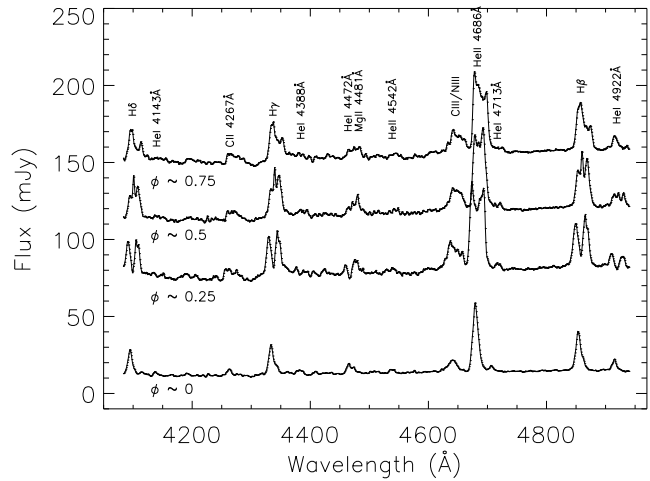}}
\noindent
\end{picture}
\vspace{75mm}
\caption{Spectra of \ipp\ taken during outburst. The spectra are
  binned into 4 orbital phase slots. An offset of 40 mJy was 
  included between consecutive spectra for display purposes.}
\label{res:avsp}
\end{figure}

The average spectra of \ipp\ during outburst shows Balmer, He\,{\sc i}
and He\,{\sc ii} as well as some Mg\,{\sc ii}, C\,{\sc ii}, C\,{\sc
  iii} and N\,{\sc iii} emission lines on a blue continuum.  The lines
appear double-peaked during most orbital phases.  Fig.~\ref{res:avsp}
shows the spectra of \ipp\ for 4 different orbital bins where the
changes of the line profiles are clear. Harlaftis (1999) presents
outburst spectra of \ipp\ at a slightly different wavelength range
where Ti\,{\sc ii} emission lines are also present.

\subsection{Doppler maps}
\label{sec:doppler}
\begin{figure*}
   \begin{picture}(100,0)(-270,250)
   \put(0,0){\includegraphics{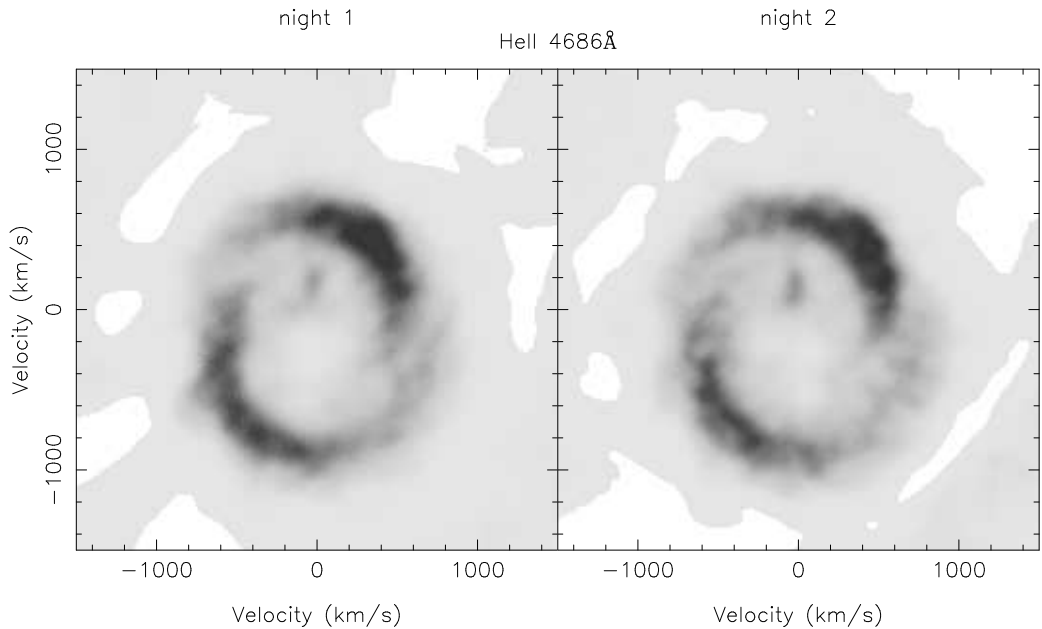}}
\noindent
\end{picture}
 \vspace{85mm}
 \caption{Doppler maps of \heii\ for both nights of observations
   obtained using the MEM mapping technique.}
\label{res:heiidopp}
\end{figure*}

We use the maximum entropy method (MEM) to construct velocity maps of
\ipp\ in different emission lines.  Fig.~\ref{res:heiidopp} shows the
images in the light of \heii, and demonstrates the presence of strong
spiral structure on each night, with little, if any, change from night
to night. There is also some emission near the secondary star. This
spiral structure has previously been observed in \ipp\ $1.5$ days and
possibly 8 days after the start of the August 1993 outburst by Steeghs
\etal (1996, 1997, 1998), and 3 days after the start of the November
1996 outburst by Harlaftis \etal (1999); the spectra taken 8 days
after the start of the August 1993 outburst only hint at the presence
of the spiral shocks.  The data presented in this paper fit in nicely
between the previous observations and have the extra advantage that,
for the first time, were taken for two consecutive nights during an
outburst. We see that the spiral structure is strong both nights and
with little alteration which indicates that the time-scale of the
duration of the spiral shocks is of the order of days instead of
hours.  As already mentioned, the data presented in this paper were
taken 5 and 6 days after the outburst had started, indicating that the
spiral shocks are long-lived structures that are probably present
throughout the entire outburst although we cannot be certain until
spectroscopy for the entire duration of an outburst is obtained.

Fig.~4 compares maps of different emission lines (\hb, \heii\ and
\hei) and also displays the data (top panels) and fits (lower panels).
To construct these maps (and those of Fig.~\ref{res:heiidopp}) the
eclipse data have been removed as the behaviour during eclipse is not
included in the model. The spiral structure is strongest in \heii\ 
whereas both \hb\ and \hei\ feature very strong emission from the
companion star. The \hei\ image is distinctly different in structure
with the upper shock stronger than the lower. This may be associated
with its weakness; for instance in computing this map we had to
simultaneously calculate one for the nearby \mgii\ line.  The upper
shock in the \hei\ maps shows a more complex structure than that of
the other lines. This structure is seen in maps from both nights,
which suggests that it is real. As for \hb, the brightest region on
the upper shock is shifted in azimuth with respect to \heii.

\begin{figure*}
\vbox to220mm{\vfil Landscape figure to go here.
\caption{}
\vfil}
\label{res:alldopp}
\end{figure*}

\subsection{Flares}

There are substantial flares in the emission lines, e.g.
in \hb\ and \heii\ at orbital phase $\sim$1.3 during the first night.

We have labelled some of the flares in the trailed spectra (top
panels) of \hb\ and \heii\ in Fig.~\ref{res:alldopp}.  The flare
events, which last for one or two spectra (i.e. a few minutes), are
clear in both lines and occur in both the blue- and red-shifted sides
of the lines.  Flares A and B occur at high radial velocities in the
blue-shifted component.  Label C is associated with a flare that took
place in the red-shifted component of the lines.

\subsection{Continuum and emission lines eclipses}

The continuum and emission line fluxes for the two nights are plotted
versus orbital phase in Fig.~\ref{res:alllc}. Orbital phases have been
calculated using Wolf et al.'s (1993) linear ephemeris:

\begin{figure*}
\begin{picture}(100,0)(10,20)
\put(0,0){\includegraphics{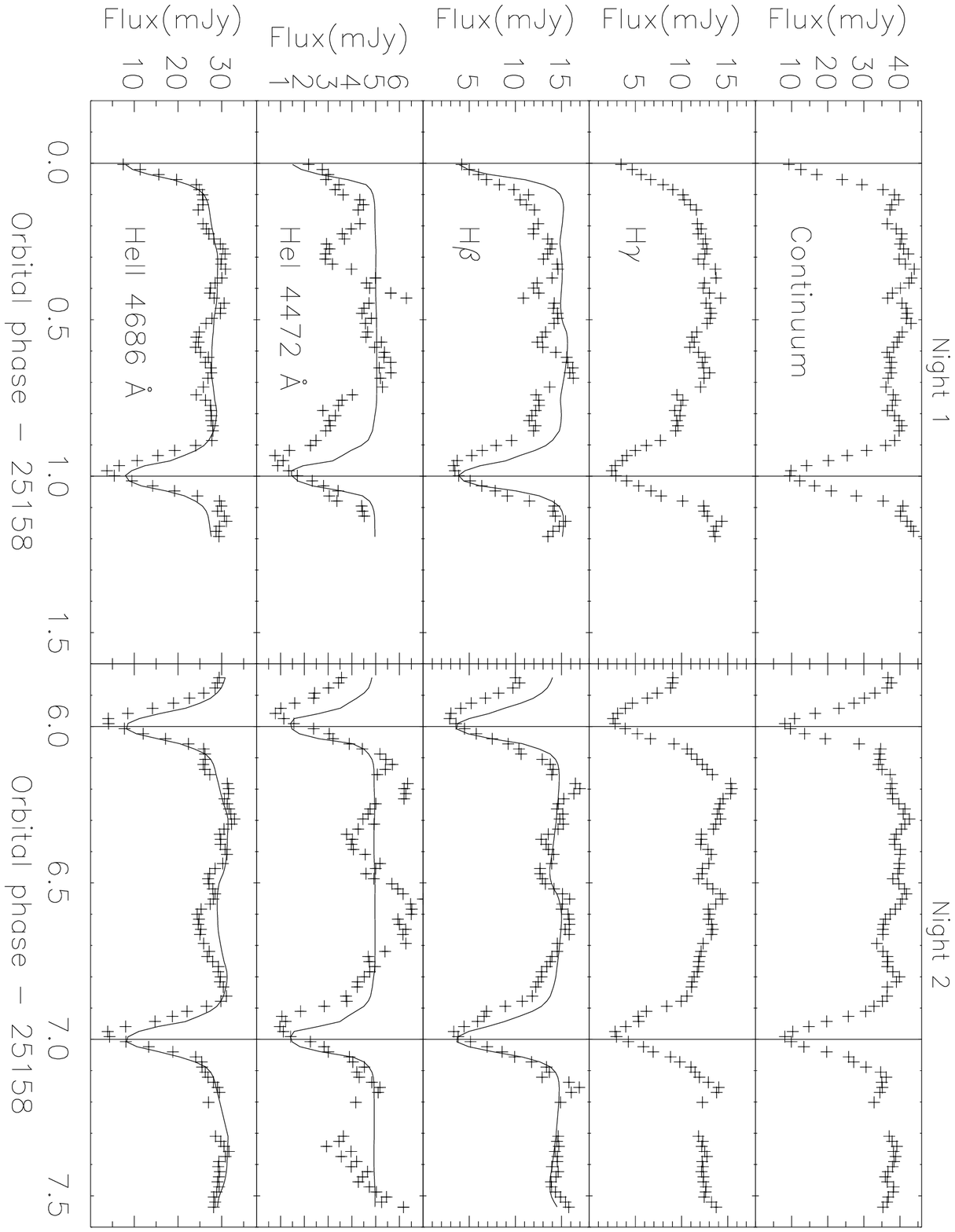}}
\noindent
\end{picture}
\vspace{97mm}
\caption{Light curves of the continuum, \hg, \hb, \hei, and \heii\
  obtained during both nights. The model light curves for \hb, \hei,
  and \heii\ are over plotted to the data.}
\label{res:alllc}
\end{figure*}

\begin{equation}
T_{0}(HJD) = 2445615.4156 + 0.15820616 E
\end{equation}

The shape of the continuum eclipses is similar to that of \heii\ 
whereas the shape of the eclipses for the other lines is quite
different. They are more asymmetric and show a slope shallower than
that of \heii\ and the continuum during both ingress and egress. The
eclipse of the emission lines is shifted to earlier-than-expected
phases; the continuum is too, but to a lesser extent.  The shift in
phase is different for different lines being the largest displacement
that of \hei\ which we singled out earlier for its asymmetry.  The
phases of mid-eclipse measured at half eclipse depth are listed in
Table~\ref{res:ecdisp} during eclipse.  Another feature of these light
curves is the variability observed between eclipses with a double-hump
in the Balmer and \hei\ lines.  This double-hump behaviour is also
seen during the September 1997 outburst in the infrared (Webb 1999)
and may result from shadowing by the spiral shocks (Steeghs priv.
comm.).  Adding the quadratic term of the ephemeris (as given by Wolf
\etal 1993) results in an orbital phase change of $\sim$0.006 which
cannot explain the shifts observed in the eclipses of the emission
lines.

\begin{table}
\begin{center}
\caption{Phases of mid-eclipse.}
\label{res:ecdisp}
\begin{tabular}{ll}
Emission line & Phase of mid-eclipse\\
\hline
\hg\ & $-$0.030 $\pm$ 0.003\\
\hb\ & $-$0.029 $\pm$ 0.004\\
\hei\ & $-$0.045 $\pm$ 0.009\\
\heii\ & $-$0.015 $\pm$ 0.001\\
Continuum & $-$0.006 $\pm$ 0.003\\
%Model & $-$0.005 $\pm$ 0.001\\
\end{tabular}
\end{center}
\end{table}

By assuming a Keplerian accretion flow we transform the velocity
coordinates of the Doppler maps to space coordinates and compute model
spectra for \hb, \heii, and \hei. These spectra, shown in the lower
panels of Fig.~4, are calculated at all orbital phases, including
during eclipse. Although the spectra during the eclipse were not used
to obtain the Doppler images, we can calculate the computed spectra
during eclipse. These spectra can be calculated from the maps by using
IP~Peg's parameters, given by Marsh \&\ Horne (1990).  The translation
from velocities to positions becomes doubtful at low velocities which
can end outside the Roche lobe. We arbitrarily decided to ignore any
eclipse of points which mapped to locations beyond r = R$_{{\rm L}1}$.
The light curves generated from the computed and real data are plotted
in Fig.~5 with solid lines and crosses respectively. The eclipses of
the computed and real data do not occur at the same time.  The model
data do not show the shift of minimum light towards earlier orbital
phases displayed by the data. Shifts of the sort seen in the data
require a large asymmetry in the emission line flux distribution about
the line of centres of the two stars. The spiral shocks do have this
asymmetry but apparently not large enough to explain the shifts that
we see. We have no plausible explanation for this.

\subsection{Companion star}

\begin{figure*}
\begin{picture}(100,0)(10,20)
\put(0,0){\includegraphics{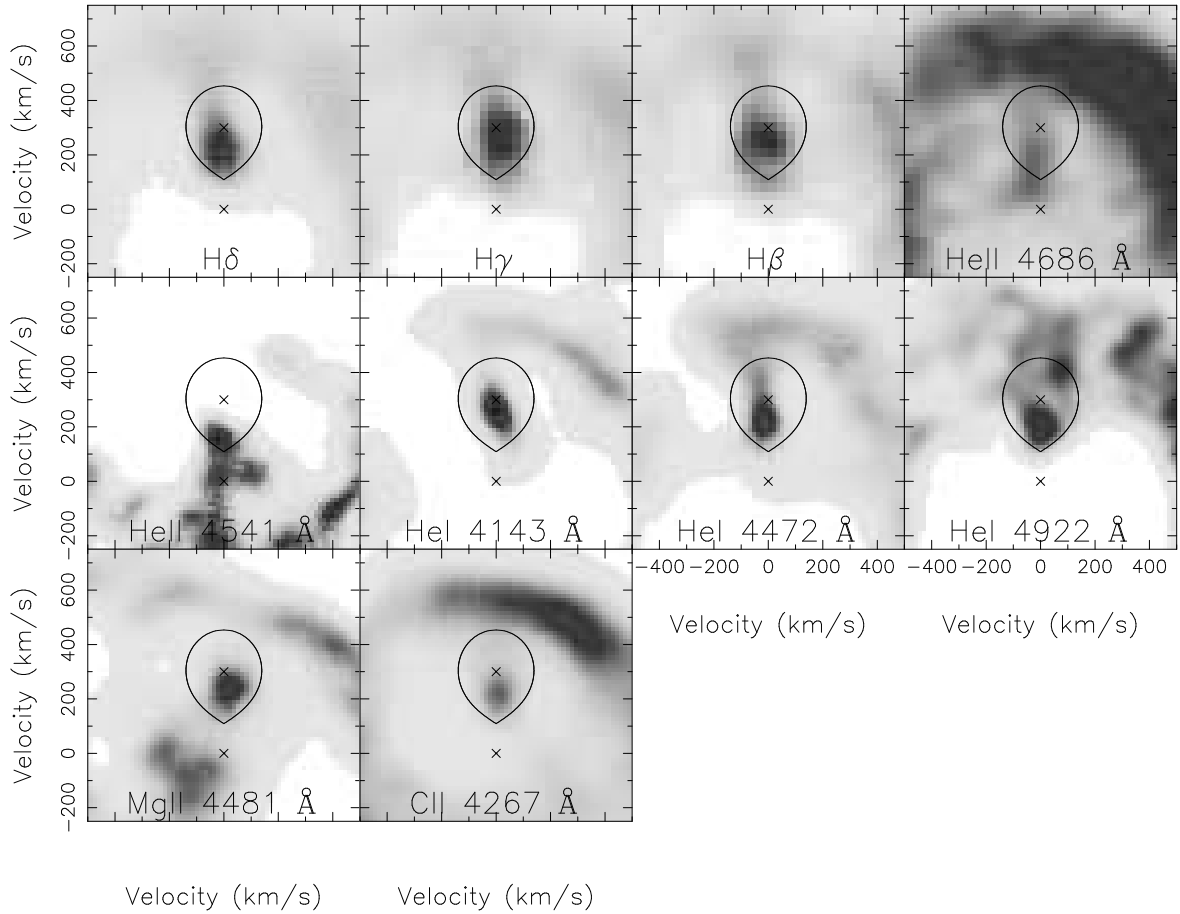}}
%\put(0,0){\special{psfile="redstarsmallsimul.ps" 
%hoffset=-192 voffset=-330 angle=270 vscale=41 hscale=41}}
\noindent
\end{picture}
\vspace{135mm}
\caption{Doppler maps in the region of the secondary star in \ipp\ for
  different emission lines.  The Roche lobe of the secondary star, and two
  crosses indicating the centre of the secondary star and the centre
  of mass of the system are also plotted.}
\label{res:comp}
\end{figure*}

We noticed in section~\ref{sec:doppler} that the secondary star
contributed to the Balmer emission and almost none of the \heii\ 
emission. In the top panels of Fig.~\ref{res:comp}, we show the
Doppler maps obtained from the combined spectra for both nights but
this time centered and expanded around the secondary star instead of
the centre of mass of the system. The symmetric component of the
accretion disc for \heii\ has been subtracted and the saturation
levels on the map adjusted to emphasize the companion star. The ratio
of the masses of both components used is $q = 0.49$ and the radial
velocity semi-amplitude of the secondary $K_2 = 300\kmsec$ (Marsh \&\ 
Horne 1990).  Maps for the Balmer lines, He\,{\sc ii}, He\,{\sc i},
Mg\,{\sc ii}, and C\,{\sc ii} are shown. We also find significant
emission coming from the red star at wavelengths: 4126.5\,\AA,
4132.5\,\AA, 4271.7\,\AA, 4383.6\,\AA, 4442.4\,\AA, 4507.8\,\AA,
4548.9\,\AA, 4823.4\,\AA, and 4890.8\,\AA. We used an LTE model of gas
combined with the atomic spectral line list of Hirata \& Horaguchi
(1995) at temperatures between 3000 and 15000 K but were unable to
identify these lines. To discard the possibility of the lines being
the result of high ionisation, we also compared them with emission
lines observed in planetary nebulae without success.

We see that the Balmer emission is more spread on the surface of the
secondary star than any of the other emission lines. This effect,
already observed in post-common envelope binaries like GD~448 (Maxted
\etal 1998), RE~1016$-$053 and RE~2013+400 (Wood, Harmer \& Lockley
1999) has been explained as the result of the Balmer lines being
intrinsically broadened due to optical depth effects; this is clear
evidence that the same effect appears in CVs. Therefore the width of
the Balmer lines is not a reliable indication of their spatial
distribution.

\begin{figure}
\begin{picture}(100,0)(-270,250)
\put(0,0){\includegraphics{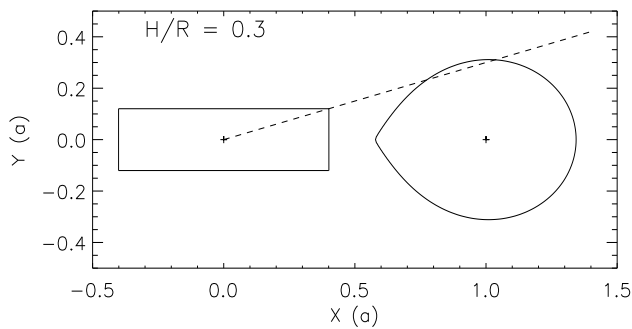}}
\noindent
\end{picture}
\vspace{55mm}
\caption{An edge-on view of  \ipp\ showing how the disc can
  shadows the equator of the secondary star. For this plot we have
  chosen $H/R=0.3$.}
\label{res:diag}
\end{figure}

The position of the irradiated region on the companion star seems to
vary from line to line but most importantly, it is located between the
inner Lagrangian point and the pole of the secondary for all emission
lines except perhaps He\,{\sc ii}. We tested for the effect of noise
by generating artificial Doppler maps with noise equivalent to that in
our data; these data showed that the noise does not cause significant
shifts in the location of the irradiated spot. The presence of an
irradiation spot on the surface of the companion star would produce a
very clear signature in its light curve, unfortunately due to the
complexity of the spectra it is difficult to extract a light curve
with only the contribution from the secondary star. However, we can
see flux modulated with the companion's motion in the trailed spectra
shown in Fig.~4 indicating the presence of such an irradiated region
near the inner Lagrangian point.  We believe that this behaviour is a
consequence of irradiation of the companion by the inner disc and
white dwarf under the influence of shadowing by the disc.
Fig.~\ref{res:diag} shows an edge-on view of \ipp\ assuming that the
disc is thick enough to shield the companion almost entirely.

\begin{figure}
\begin{picture}(100,0)(-270,250)
\put(0,0){\includegraphics{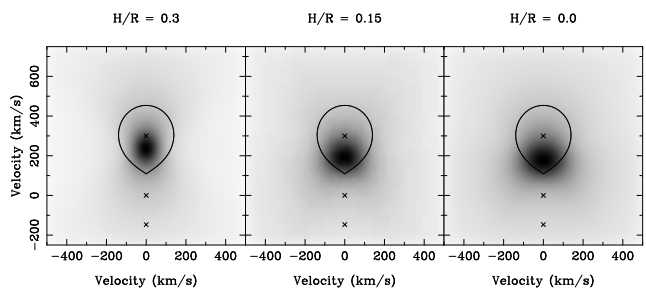}}
\noindent
\end{picture}
\vspace{45mm}
\caption{Simulated irradiation pattern on the surface of the red star
  for three different disc thicknesses, $H/R = 0.0$ (left), $0.15$
  (centre), and $0.30$ (right). The Roche lobe of the companion star
  is also plotted in all maps to help identify the position of the
  irradiated region. The centre of mass and the position of the white
  dwarf are marked with a cross.}
\label{res:compsim}
\end{figure}

Only a region near the poles of the companion star can see the inner
disc and white dwarf, and therefore we expect to see emission
displaced from the L1 point towards the centre of mass of the
secondary star. This displacement survives the process of Doppler
tomography, even though it does not strictly satisfy the assumptions
which underly it: in Fig.~\ref{res:compsim} we show maps derived
taking shadowing and Roche lobe geometry into account for $H/R = 0.0$,
$0.15$, and $0.30$ (anything larger than 0.31 prevents all light from
reaching the secondary star).  Comparing this figure with the real
maps, suggests that the disc has an effective $H/R \approx 0.2$. The
$H/R$ may be able to vary owing to wavelength-dependent opacity at the
EUV wavelengths appropriate for photoionisation of the various species
involved.  For instance, the Balmer lines require photons with $h \nu
> 13.6\,{\rm eV}$, compared to $24\,{\rm eV}$ for \hei\ and $54\,{\rm
  eV}$ for \heii. We would expect the opacity at $13.6\,{\rm eV}$ to
be largest (because of neutral hydrogen), and thus the disc to appear
thickest for the Balmer lines. Our data are suggestive of this effect,
although not conclusive. Higher resolution observations would be of
use because apart from the Balmer emission, the irradiated features
are not resolved in our data.

\section{Conclusions}

We have discovered spiral shocks in the August 1994 outburst of the
dwarf nova \ipp\ 5 and 6 days after the system went into outburst and
with no noticeable diminution in strength from one day to the next.
Our data extend the duration over which spiral structure is known to
remain strong, and suggest that they could last the entire outburst.
The eclipses of the emission lines are shifted by about 0.015--0.045
towards earlier phases than that of the white dwarf. These shifts
towards earlier phases require a large-scale asymmetry about the line
of centres of the two stars larger than the spiral shocks suggest. We
cannot explain this. The disc is also apparently variable on short
timescales as we see several strong flares in both Balmer and He\,{\sc
  ii} emission lines.

Doppler maps of line emission on the companion star show that it is
located between the inner Lagrangian point and the poles of the star,
and that it avoids the equator. This is strong evidence for shielding
of the companion star by the disc. From comparison with models, the
shielding must be substantial, requiring a height-to-radius ratio in
the disc of $H/R \approx 0.2$. There is some evidence for variability
in the location of the emission, as might be expected if the disc
appears to vary in thickness according to the threshold wavelength
needed to drive the respective emission line. The Balmer emission from
the companion is clearly broader than the other emission lines,
supporting evidence from detached binaries for non-Doppler broadening
in this line.

\subsection*{Acknowledgments}

The Isaac Newton Telescope is operated on the island of La Palma by
the Isaac Newton Group in the Spanish Observatorio del Roque de los
Muchachos of the Instituto de Astrof\'{\i}sica de Canarias. In this
research, we have used, and acknowledge with thanks, data from the
AAVSO International Database, based on observations submitted to the
AAVSO by variable star observers worldwide. The reduction and analysis
of the data were carried out on the Southampton node of the STARLINK
network. LM-R wishes to thank L. Gonz{\' a}lez Hern{\' a}ndez for
computer support.

% \bibitem{1} Bobinger\, A., Horne\, K., Mantel\, K.\,H., Wolf\, S.,
%   1997, A\&A, 327, 1023-1038
% \bibitem{2} Harlaftis\, E.\,T., Marsh\, T.\,R., Dhillon\, V.\,S.,
%   Charles\, P.\,A., 1994, MNRAS, 267, 473-480
% \bibitem{16} Horne\, K., 1992, 12th North American Workshop on CVs and
%  LMXB, San Diego University Press.
% \bibitem{17} Kwee\, K.\,K., van Woerden\, H., 1956,
%   Bull. Astron. Inst. Neth. 12, 327
% \bibitem{3} Marsh\, T.\,R., 1988, MNRAS, 231, 1117-1138
% \bibitem{5} Martin\, J.\,S., Friend\, M.\,T., Smith\, R.\,C., Jones\,
%   D.\,H.\,P., 1989, MNRAS, 240, 529-531
% \bibitem{6} Piche\, F., Szkody\, P., 1989, AJ, 98, 6
% \bibitem{9} Steeghs\, D., Stehle\, R., 1998, MNRAS, submitted
% \bibitem{10} Wolf\, S., Barwing\, H., Bobinger\, A., Mantel\, K.\,H.,
%   Simic\, D., 1998, A\&A, 332, 984-998
% \bibitem{12} Wood\, J., Crawford\, C.\, S., 1986, MNRAS, 222, 645-654

\end{document}